\documentclass{article}
\usepackage{graphicx}
\usepackage{booktabs} 
\usepackage{amsmath}
\usepackage{amsfonts}
\usepackage{authblk}
\usepackage{xcolor}
\usepackage{hyperref}
\usepackage{array}
\usepackage{multirow} 
\usepackage{arydshln}
\usepackage[switch]{lineno}
\usepackage{makecell}
\usepackage[numbers,sort&compress]{natbib}

\newcommand{\tabref}[1]{(Table~\ref{#1})}

\newcommand{\figref}[2]{(Fig.~\ref{#1}#2)}

\newcommand{\ie}{i.e.,~}
\newcommand{\eg}{e.g.,~}
\setcitestyle{numbers,square,super,comma,compress}

\usepackage[accepted]{icml2021}

\icmltitlerunning{How Evaluation Choices Distort the Outcome of Generative Drug Discovery}

\begin{document}
\twocolumn[
\icmltitle{How Evaluation Choices Distort the Outcome of Generative Drug Discovery}

\begin{icmlauthorlist}
\icmlauthor{Rıza Özçelik}{tue,clt}
\icmlauthor{Francesca Grisoni*}{tue,clt}
\end{icmlauthorlist}

\icmlaffiliation{tue}{Institute for Complex Molecular Systems and Dept. Biomedical Engineering, Eindhoven University of Technology, Eindhoven, The Netherlands.}
\icmlaffiliation{clt}{Centre for Living Technologies, Alliance TU/e, WUR, UU, UMC Utrecht, The Netherlands}

\icmlcorrespondingauthor{Francesca Grisoni}{f.grisoni@tue.nl}

\vskip 0.3in
]

\newcommand{\beginsupplement}{%
    \setcounter{table}{0}
    \renewcommand{\thetable}{S\arabic{table}}%
    \setcounter{figure}{0}
    \renewcommand{\thefigure}{S\arabic{figure}}%
 }

\printAffiliationsAndNotice{}  
\begin{abstract}
\noindent ``How to evaluate the de novo designs proposed by a generative model?" Despite the transformative potential of generative deep learning in drug discovery, this seemingly simple question has no clear answer. The absence of standardized guidelines challenges both the benchmarking of generative approaches and the selection of molecules for prospective studies.  In this work, we take a fresh -- \textit{critical} and \textit{constructive} -- perspective on de novo design evaluation. By training chemical language models, we analyze approximately 1 billion molecule designs and discover principles consistent across different neural networks and datasets. We uncover a key confounder: the size of the generated molecular library significantly impacts evaluation outcomes, often leading to misleading model comparisons. We find increasing the number of designs as a remedy and propose new and compute-efficient metrics to compute at large-scale. We also identify critical pitfalls in commonly used metrics -- such as uniqueness and distributional similarity -- that can distort assessments of generative performance. To address these issues, we propose new and refined strategies for reliable model comparison and design evaluation. Furthermore, when examining molecule selection and sampling strategies, our findings reveal the constraints to diversify the generated libraries and draw new parallels and distinctions between deep learning and drug discovery. We anticipate our findings to help reshape evaluation pipelines in generative drug discovery, paving the way for more reliable and reproducible generative modeling approaches.

\textit{keywords:} evaluation, de novo design, deep learning, small molecules, generative modeling.
\end{abstract}
\section*{Introduction}
Discovering new therapeutics is an adventure as old as human civilization. However, finding new drug molecules is more resource-intensive today than ever \cite{wouters2020estimated,dimasi2016innovation}. A key challenge lies in the vastness of the `chemical universe', which is estimated to contain more than $10^{60}$ drug-like molecules where compounds with desirable biological properties are exceedingly rare  \cite{bohacek1996art}. Artificial intelligence (AI) has emerged as a transformative technology for drug discovery, to help find the `needle in the haystack'. By supporting virtual screening \cite{stokes2020deep,liu2023deep,wong2024discovery} and de novo molecule design \cite{godinez2022design,wan2022deep,moret2023leveraging,li2022generative,isigkeit2024automated,xia2024target}, AI can narrow down the chemical universe, and it is nowadays widely adopted in academia and industry \cite{bian2021generative,gangwal2024unlocking,cheng2021molecular,volkamer2023machine,catacutan2024machine}. Generative deep learning has garnered particular attention for drug discovery. Powered by deep neural networks, these models can learn how to generate molecules with desired properties on demand, and have already demonstrated success in prospective studies \cite{yuan2017chemical,merk2018novo,grisoni2021combining,godinez2022design,ballarotto2023novo,du2024machine}.

Generative drug discovery generally involves three stages: \textit{train}, \textit{generate}, and \textit{evaluate}. After almost a decade from its initial introduction \cite{segler2018generating,gomez2018automatic}, prolific research has standardized many aspects of model  \textit{training}\cite{bian2021generative,sousa2021generative, gangwal2024unlocking,cheng2021molecular,grisoni2023chemical,ozccelik2024chemical} and molecule \textit{generation} \cite{gupta2018generative,moret2023leveraging,ozccelik2024chemical,sousa2021generative}. However, the third stage -- \textit{evaluation} -- remains relatively underinvestigated, with choices left to the single practitioners. The \textit{evaluation} of molecular designs (\eg in terms of their overall quality, relevance, and ultimately, ranking) holds a crucial role. First, selecting the best candidates from thousands of designs determines the success or failure of follow-up experiments. Second, robust evaluation of the generated molecules is essential to monitor progress in the field and to compare different approaches. Yes, despite notable efforts to standardize model evaluation\cite{polykovskiy2020moses,brown2019guacamol,arus2019exploring,nie2024durian,thomas2024molscore}, no consensus within the community has been reached \cite{renz2019failure,bender2022evaluation,sousa2021generative,martinelli2022generative}.

Here, we dive into design evaluation, with a \textit{critical} and \textit{constructive} perspective. We conduct a systematic study using chemical language models -- a widely applied and experimentally validated family of generative approaches\cite{du2024machine}. By capitalizing on their scalability, we generate and evaluate 10$^9$ de novo designs across three state-of-the-art architectures and four datasets. Our results uncover a previously overlooked pitfall: The size of a design library can systematically bias the evaluation, and at times even falsify the scientific findings by overshadowing molecular quality. We find that increasing the number of designs helps avoiding this pitfall, and propose new, scalable evaluation metrics. We then uncover the risks of relying on design frequencies, a common criterion for molecule selection, and develop solutions to mitigate the risks. We next leverage the tools we develop to dig deeper into the \textit{generation} stage, and expose inherent constraints to achieve high design diversity. Finally, we distill our findings into concrete challenges, methodological improvements, and practical strategies to enhance generative model evaluation. By addressing these critical aspects, we aim to advance how generative models are assessed and how molecules are selected for prospective studies in drug discovery.

\section*{Results and Discussion}

While many approaches exist to design molecules de novo\cite{lin_diffbp_2022,cremer_flowr_2025,noauthor_evaluation_nodate,peng_moldiff_2023,vignac_digress_2023,liu_graph_2024,shi_graphaf_2020,verma_modular_2022,luo_graphdf_2021,zang_moflow_2020}, `chemical language' models (CLMs) have been among the most successful\cite{grisoni2023chemical,skinnider2021chemical,flam2022language}. CLMs are trained to generate molecules in the form of molecular strings, such as Simplified Molecular Input Line Entry Systems (SMILES)\cite{weininger1988smiles} and Self-referencing embedded strings (SELFIES)\cite{krenn2022selfies}. Since CLMs are the most widely used approach for molecule design in practice\cite{du2024machine}, and
can be trained and used to generate molecules in a time-efficient manner \cite{ozccelik2024chemical}, they constitute an ideal choice for a large-scale analysis like ours.

Here, we use three deep CLM architectures: (i) Recurrent neural networks with long short-term memory cells, (LSTM)\cite{hochreiter1997long,segler2018generating}, which learn from and generate chemical sequences one symbol (`token') at a time; (ii) Generative Pretrained Transformers (GPT) \cite{radford2019language,bagal2021molgpt}, which, via the attention mechanism\cite{bahdanau2015neural} learn all pair relationships between input tokens; and (iii) Structured State-Space Sequence models (S4) \cite{gu2022efficiently,ozccelik2024chemical}, which were recently introduced, and learn from entire sequence at once, while generating token-by-token. 
After pre-training the CLMs on 1.5M canonical SMILES strings from ChEMBLv33\cite{gaulton2017chembl}, they were fine-tuned on bioactive molecules of three macromolecular targets relevant for drug discovery \cite{sun2017excape}: (a) Dopamine Receptor D3 (DRD3), (b) Peptidyl-prolyl cis/trans Isomerase NIMA-interacting 1 (PIN1), and Vitamin D Receptor (VDR). DRD3, a G protein-coupled receptor, is used to study neuropsychiatric\cite{sokoloff2006dopamine} and PIN1 is an enzyme that regulates multiple cancer-driving pathways\cite{zhou2016isomerase}, while VDR, a nuclear receptor, is studied to prevent cancer progression \cite{feldman2014role}. Together, these targets represent three protein families and cover a broad spectrum of therapeutic areas, making them suitable benchmarks for generative modeling in drug discovery.

The fine-tuning was repeated five times for each target, with a different random set of 320 bioactive molecules each. From each fine-tuned CLM, we sampled 1,000,000 molecules in the form of SMILES strings (using multinomial sampling, Equation (\ref{eq:t_sampling})). The described pipeline aligns with the popular transfer learning strategies for de novo design \cite{segler2018generating,cai2020transfer,brown2019guacamol}.

\subsection*{Too few generated designs cause misleading findings}

``How many designs should I generate?" Every de novo design study faces this question. Although an arbitrary number of SMILES strings could be generated, 1000 and 10,000 designs are typical choices for model evaluation\cite{polykovskiy2020moses}.
However, since generative molecule design involves sampling from a learned probability distribution, a minimum library size may be required to ensure a representative overview of the model's output.
Here, we aim to shed light on (a) what library size is sufficient to evaluate the quality of designs comprehensively, and (b) whether the chosen number of designs affects the evaluation outcomes. To this end, we evaluated the following aspects in an increasing number of de novo designs:
\begin{itemize}
    \item \textit{Similarity between de novo designs and fine-tuning sets}. We measured the Frech\'et ChemNet Distance (FCD) \cite{preuer2018frechet} between the designs and the fine-tuning molecules, which captures the biological and chemical similarity of two molecular sets (through the ChemNet \cite{mayr2018large} model). Moreover, we computed the Frech\'et distance \cite{frechet1957distance} on five molecular descriptors (Frech\'et Descriptor Distance, FDD, \textit{see} Methods) of the designs to the fine-tuning compounds. The lower the FDD, the closer the designs and fine-tuning molecules are in terms of the distribution of their physicochemical properties. As controls, we computed the FCD and FDD values of 128 held-out actives and 1280 inactives for each data split, with the hypothesis that active molecules should be closer to the fine-tuning set than the inactive ones.
    
    \item \textit{Internal diversity of designed libraries.} We calculated three metrics: (a) uniqueness, that is, the fraction of unique (and `chemically' valid) canonical SMILES strings generated, (b) the number of clusters containing structurally distant molecules as identified via sphere exclusion algorithm \cite{xiemuch,renz2019failure} (related to `\#Circles' \cite{xiemuch,renz2019failure}), and (c) number of unique substructures, identified via Morgan algorithm \cite{rogers2010extended}. To our knowledge, this is the first study that uses this latter metric to evaluate internal diversity.
    
\end{itemize}
\begin{figure*}[t]
    \centering
    \includegraphics[width=\textwidth]{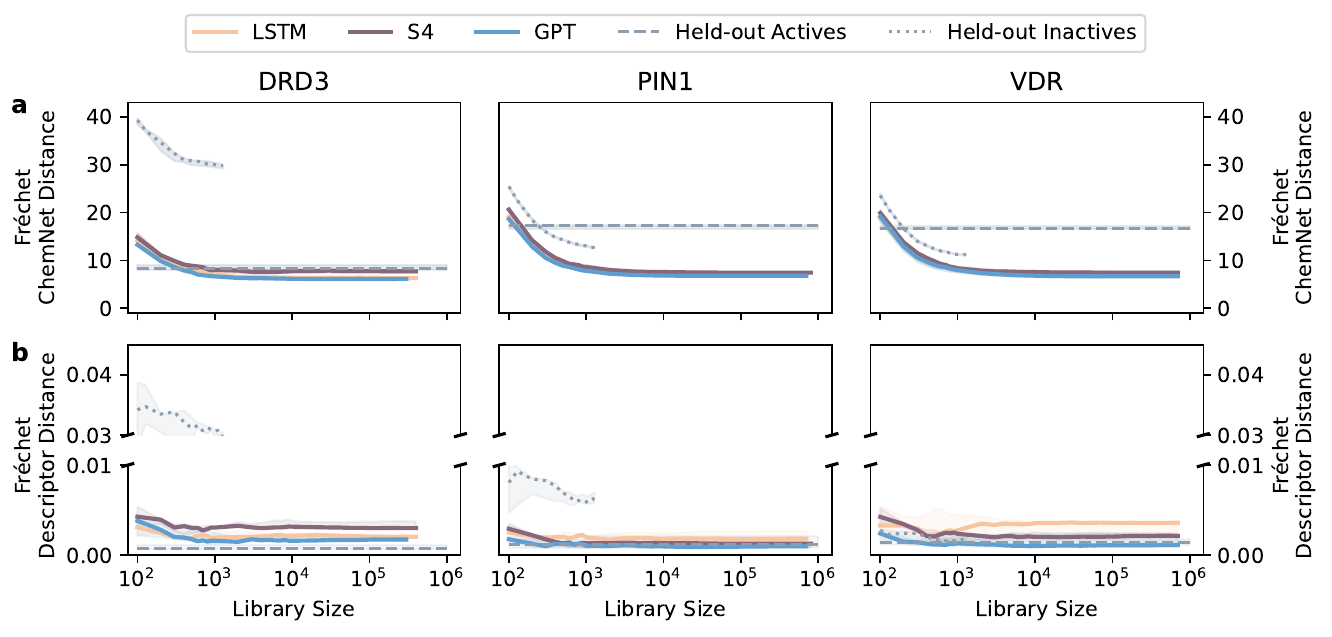}
    \caption{\textit{Number of de novo designs as a key confounder - similarity to existing molecules.} Frech\'et ChemNet Distance (FCD, \textbf{a}) and Frech\'et Descriptor Distance (FDD, \textbf{b}) are measured in increasing library sizes. Solid lines denote the median distance between design libraries and respective fine-tuning sets, across five repetitions ($n=5$), and shaded regions display the first and third quartiles. Dashed lines display the median distance of held-out actives ($n=128$) and inactives (100 $\leq n \leq 1280$), to the training sets.
    }
    \label{fig:size_trap_similarity}
\end{figure*}

\begin{figure*}[t]
    \centering
    \includegraphics[width=\textwidth]{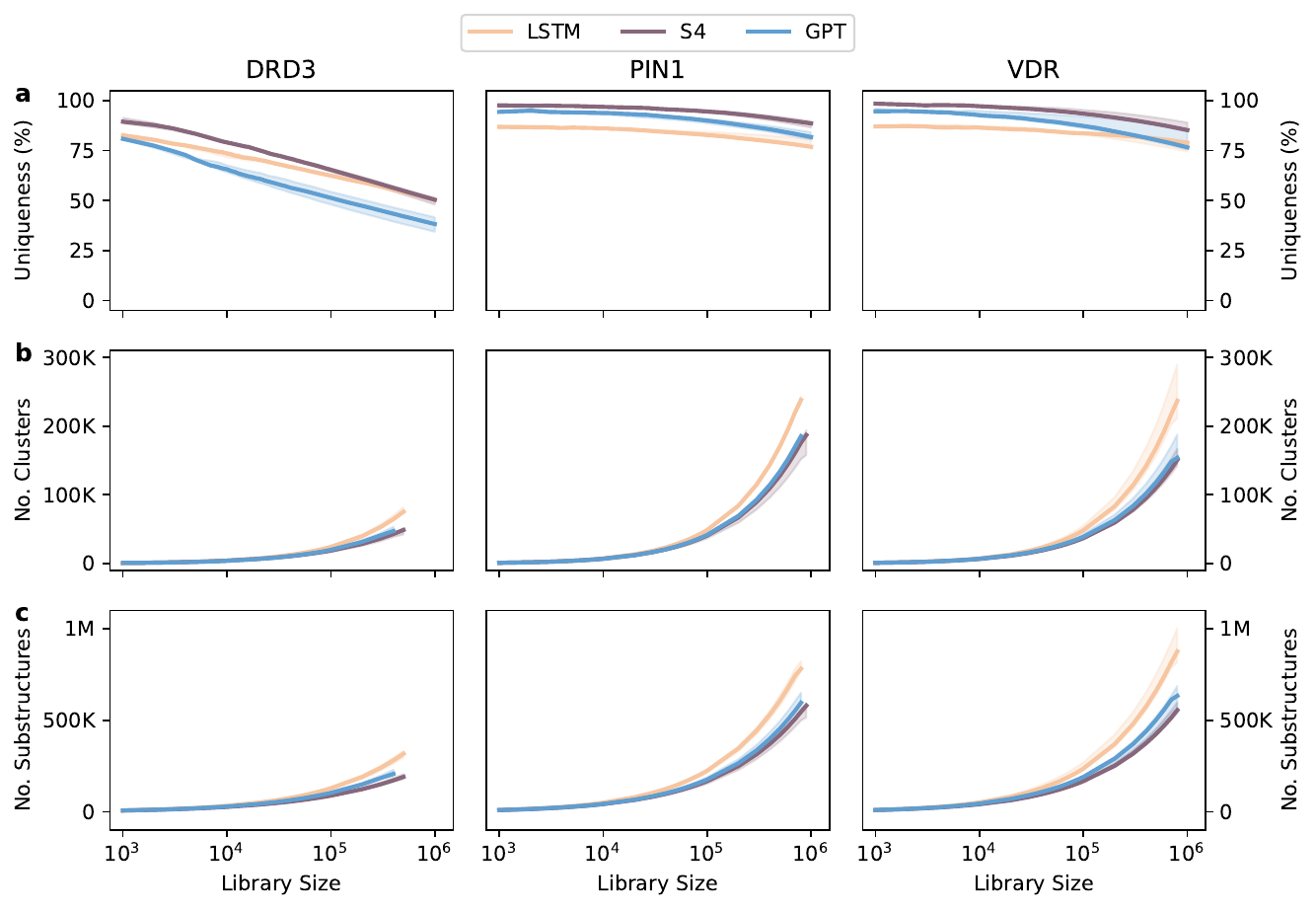}
    \caption{\textit{Number of de novo designs as a key confounder - internal diversity.} Three internal diversity metrics are measured in increasing library sizes. \textbf{(a)} Uniqueness, that is, the fraction of distinct designs among the chemically-valid ones. \textbf{(b)} Number of clusters, computed via sphere exclusion clustering, denotes the number of structurally distant molecules in the library. \textbf{(c)} Number of substructures, \ie number of unique Morgan keys \cite{rogers2010extended}. For all figures, lines display the median score measured across five fine-tuning repetitions ($n$=5) and the shaded regions show the first and third quartiles.
    }
    \label{fig:size_trap_diversity}
\end{figure*}

We systematically evaluated these aspects by varying the size of the generated library of de novo designs, from 10$^2$ to 10$^6$ molecules.

\underline{Similarity}. A relationship was observed between distribution similarity and the number of de novo designs considered \figref{fig:size_trap_similarity}{}. FCD to fine-tuning molecules decreased across all targets when increasing the library size \figref{fig:size_trap_similarity}{a}, reaching a plateau. Such an FCD plateau was systematically reached when more than 10,000 designs were considered -- a higher number than what is usually considered in de novo design studies. 
Although the authors of FCD recommend using at least 5000 molecules in each set to be compared \cite{preuer2018frechet}, our analysis shows that FCD can also be used with smaller training set sizes typical of drug discovery campaigns, since FCD values converge when enough designs are generated.

\begin{figure*}[h]
    \centering
    \includegraphics[width=\textwidth]{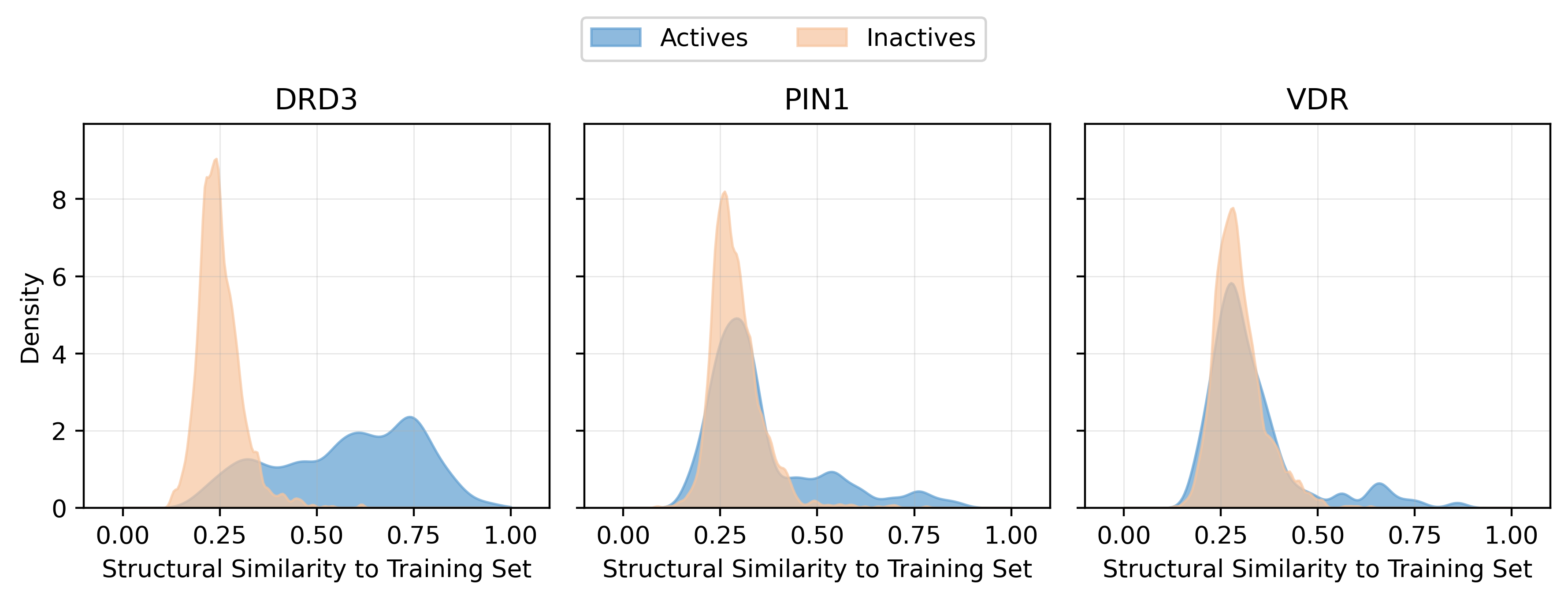}
    \caption{\textit{Structural similarity of held-out active and inactive molecules to fine-tuning sets.} The structural similarity is quantified as Tanimoto similarity between Morgan fingerprints (\texttt{radius=2}, \texttt{nBits=2048}). The maximum similarity of each held-out molecule to the fine-tuning set is computed and visualized as a distribution.}
    \label{fig:train_test_sim}
\end{figure*}

The FCD between inactive and fine-tuning molecules was, contrary to expectations, lower than that of the active molecules that were held out for PIN1 and VDR \figref{fig:size_trap_similarity}{a}. This discrepancy is because the number of inactives is nine times greater than the number of actives. DRD3 forms an exception here, due to the higher structural similarity between its held-out actives and training set \figref{fig:train_test_sim}{}. The design libraries reached FCD values lower than those of the held-out actives across proteins, again as an effect of the library size.  These findings demonstrate the cruciality of using the same number of molecules when comparing molecule libraries via FCD.

\begin{figure*}[h]
    \centering
    \includegraphics[width=\textwidth]{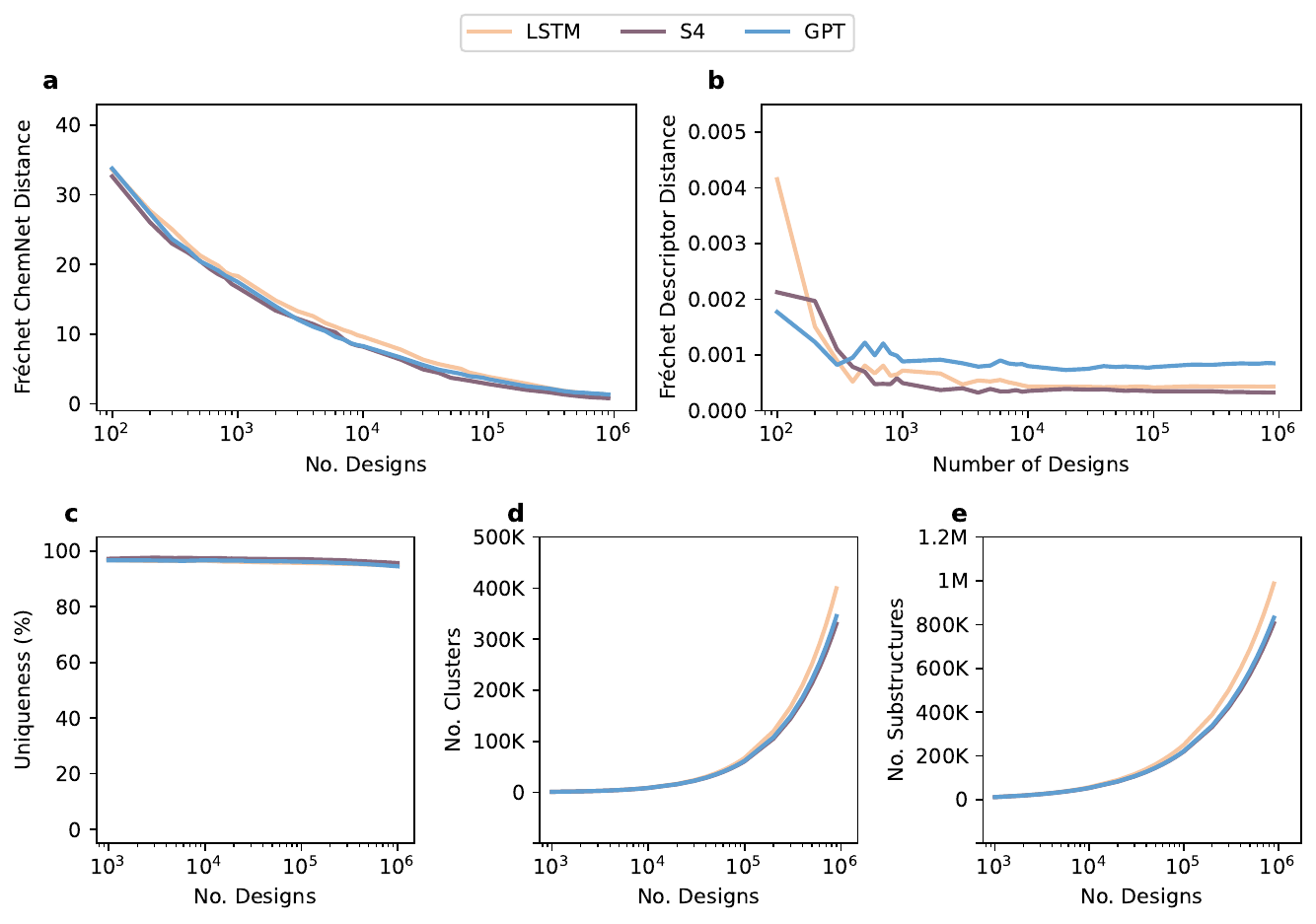}
    \caption{\textit{Pretraining designs similarity and diversity.}1,000,000 designs are generated via each pretrained model. FCD \textbf{(a)}, FDD \textbf{(b)}, uniqueness \textbf{(c)}, number of clusters \textbf{(d}), and number of structures \textbf{(e)} are reported in increasing number of designs.
    }
    \label{fig:pt_sim_div}
\end{figure*}

 Measuring FCD of the pretrained model designs to the pretraining set demonstrates the same behavior, albeit convergence required over 1,000,000 designs \figref{fig:pt_sim_div}{a}. This might be due to the high internal diversity of the pretraining set, which requires a higher number of designs to mimic with a design library. Such a `late’ convergence further underscores the importance of reporting trends in FCD values, rather than a single FCD score as in current benchmarks.

Unlike FCD, FDD scores held-out actives as more similar to the training set than inactives, across targets and scales \figref{fig:size_trap_similarity}{b}. However, FDD also decreases as the library size increases, revealing that it is also sensitive to the number of molecules used to compute the distance. The same pattern emerges also for the designs generated by the pretrained models \figref{fig:pt_sim_div}{b}, with FDD values converging similarly (unlike what was previously observed for FCD. Together, our findings underscore the library size, an overlooked parameter of the evaluation stage, as a key confounding factor of measured distributional distance between molecule libraries. We term this confounding effect `size trap'.

\begin{figure*}[h]
    \centering
    \includegraphics[width=\textwidth]{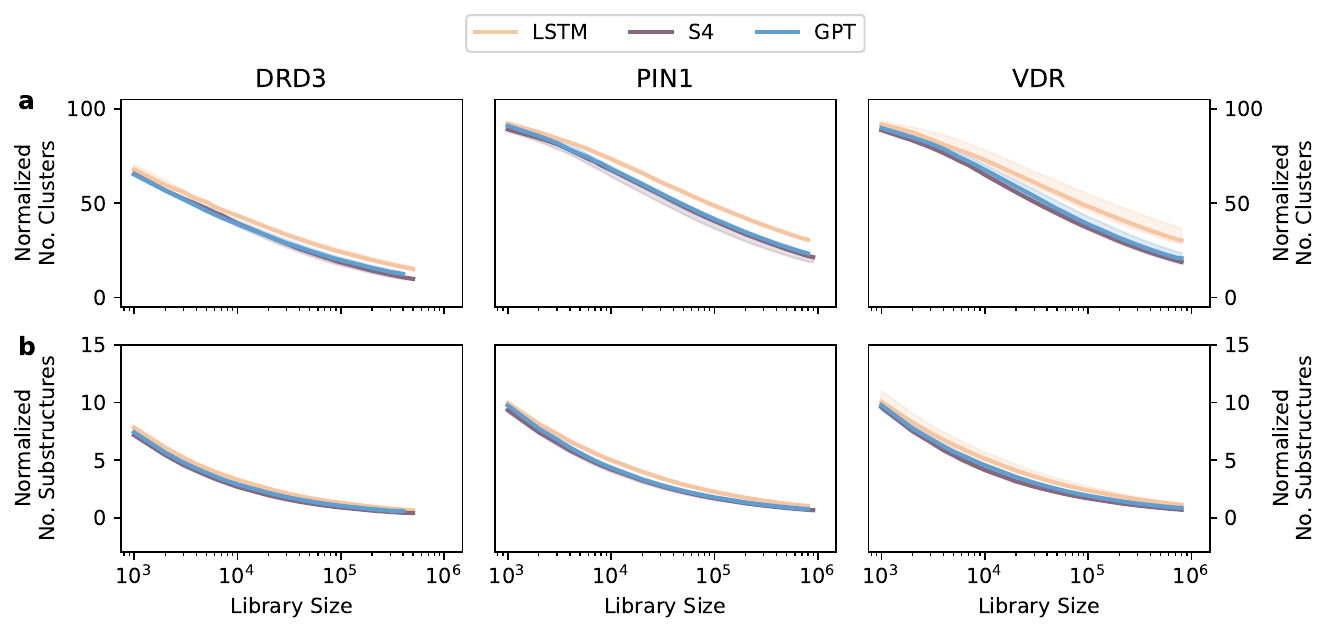}
    \caption{\textit{Diversity metrics divided by library size}. Number of substructures and clusters are computed in increasing library sizes and divided by the number of molecules in the library. The same experimental and visualization setups are used as Figure \ref{fig:size_trap_diversity}.}
    \label{fig:size_trap_div_norm}
\end{figure*}

\begin{figure*}
    \centering
    \includegraphics[width=\textwidth]{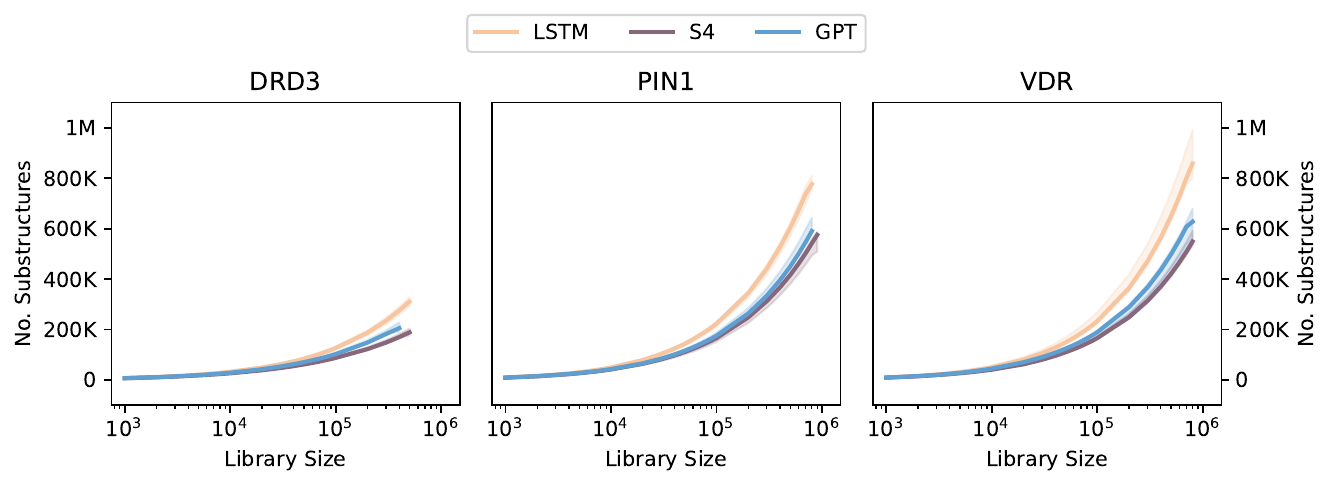}
    \caption{\textit{Effect of synthesizability filtering on internal diversity.} Synthetic accessibility of the designs is computed \cite{ertl2009estimation}and the generations with a score above six are filtered out \cite{ertl2009estimation}. Number of substructures is computed with the remaining compounds. The same experimental and visualization setups are used as Figure \ref{fig:size_trap_diversity}.}
    \label{fig:size_trap_div_sa}
\end{figure*}

\begin{figure*}
    \centering
    \includegraphics[width=.7\textwidth]{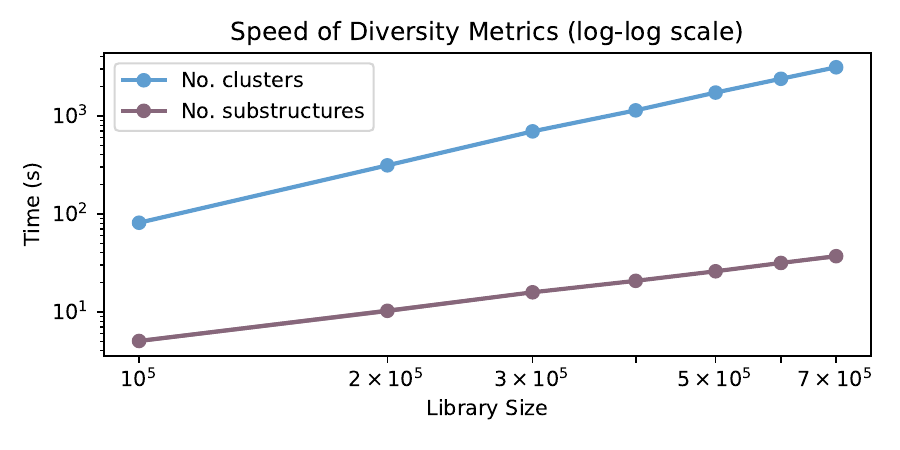}
    \caption{\textit{Speed comparison of internal diversity metrics.} Number of clusters and substructures are computed in increasing library sizes, and the number of seconds per computation is measured. Each computation is repeated 10 times, and the average time passed is reported on a log-log scale.}
    \label{fig:n_clusters_subs_speed}
\end{figure*}

\underline{Internal Diversity.} 
Uniqueness -- commonly reported to compare generative approaches -- decreases with increasing number of designs and can lead to ranking models differently depending on the generated library size \figref{fig:size_trap_diversity}{a}. The number of clusters also depends on the size of the library, with performance differences becoming progressively more pronounced for larger libraries \figref{fig:size_trap_diversity}{b}. Unlike uniqueness, the models' relative performance remains consistent across scales when the number of clusters is considered. Thus, we view uniqueness as a `sanity check' for mode collapse, rather than a reliable diversity metric. Finally, the number of substructures shows the same trend as the number of clusters (Pearson correlation coefficient larger than 99\% across experiments), also when these scores are divided by the library size \figref{fig:size_trap_div_norm}{\text{}}, and when non-synthesizable designs were filtered out before computation \figref{fig:size_trap_div_sa}{\text{}}.
Finally, all internal diversity metrics display similar trends before and after fine-tuning \figref{fig:pt_sim_div}{c-e}, while the number of substructures is up to 85 times faster to compute \figref{fig:n_clusters_subs_speed}{\text{}} than number of clusters, making it a robust and fast alternative to assess internal diversity at large scale.

\begin{figure*}[t]
    \centering
    \includegraphics[width=\textwidth]{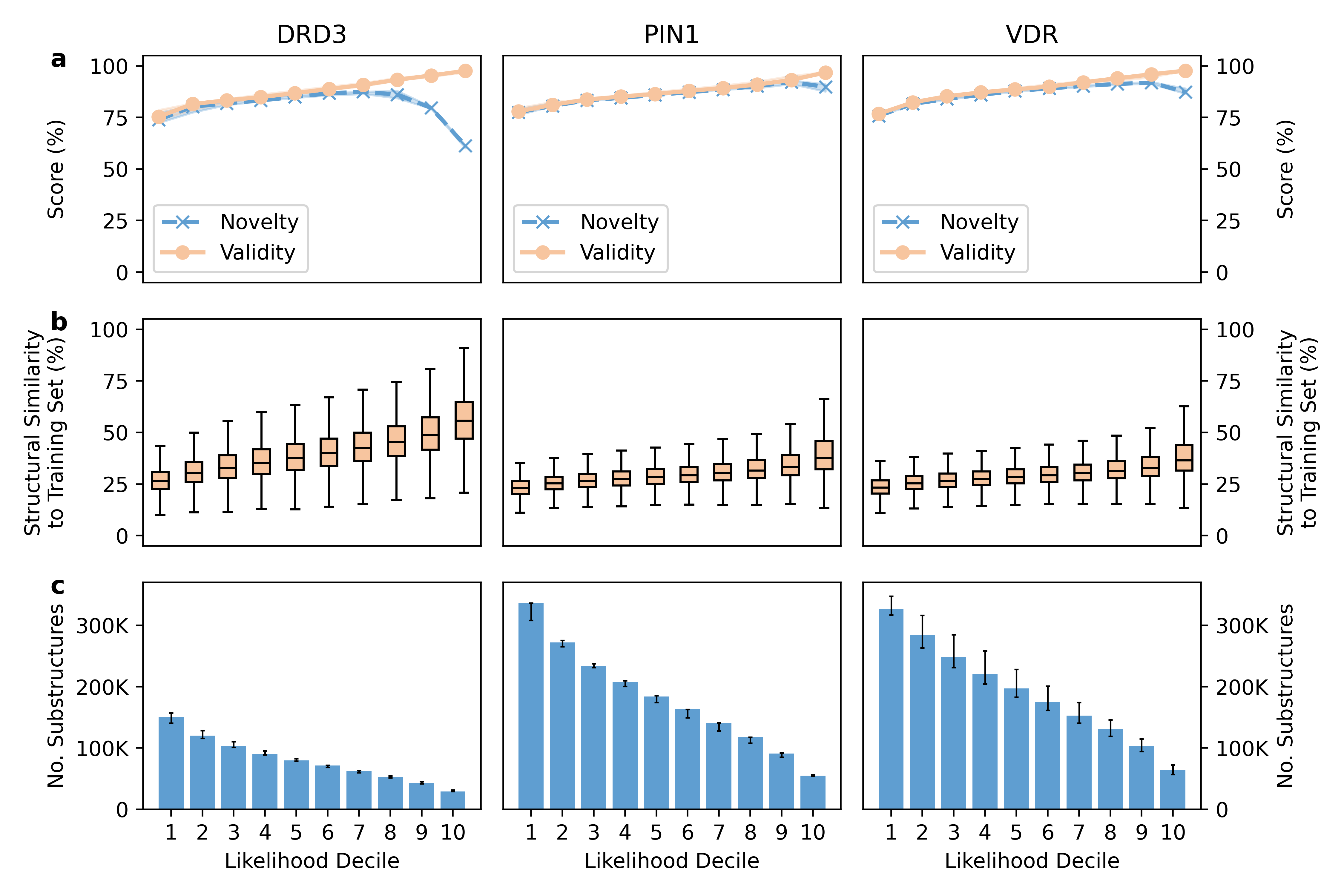}
    \caption{\textit{Navigating large design libraries.} We bin the designs per protein target into ten increasing likelihood bins and compute metrics for the designs in each decile. \textbf{(a)} Fraction of valid (validity) and unique molecules not in the respective training set (novelty) are computed. The lines represent the median across five fine-tuning campaigns, and the shaded regions mark the first and third quartiles. 
    \textbf{(b)} Structural similarity of the designs to the training set per decile is computed via Tanimoto similarity over extended connectivity fingerprints \cite{rogers2010extended}. Similarities are pooled across five repetitions and visualized as a box plot.
    \textbf{(c)} The diversity in each decile is computed via the number of substructures. Bar heights denote the median across runs, while the error bars mark the first and third quartiles. 
    }
    \label{fig:explore_vs_exploit}
\end{figure*}

\begin{figure*}[t]
    \centering
    \includegraphics[width=\textwidth]{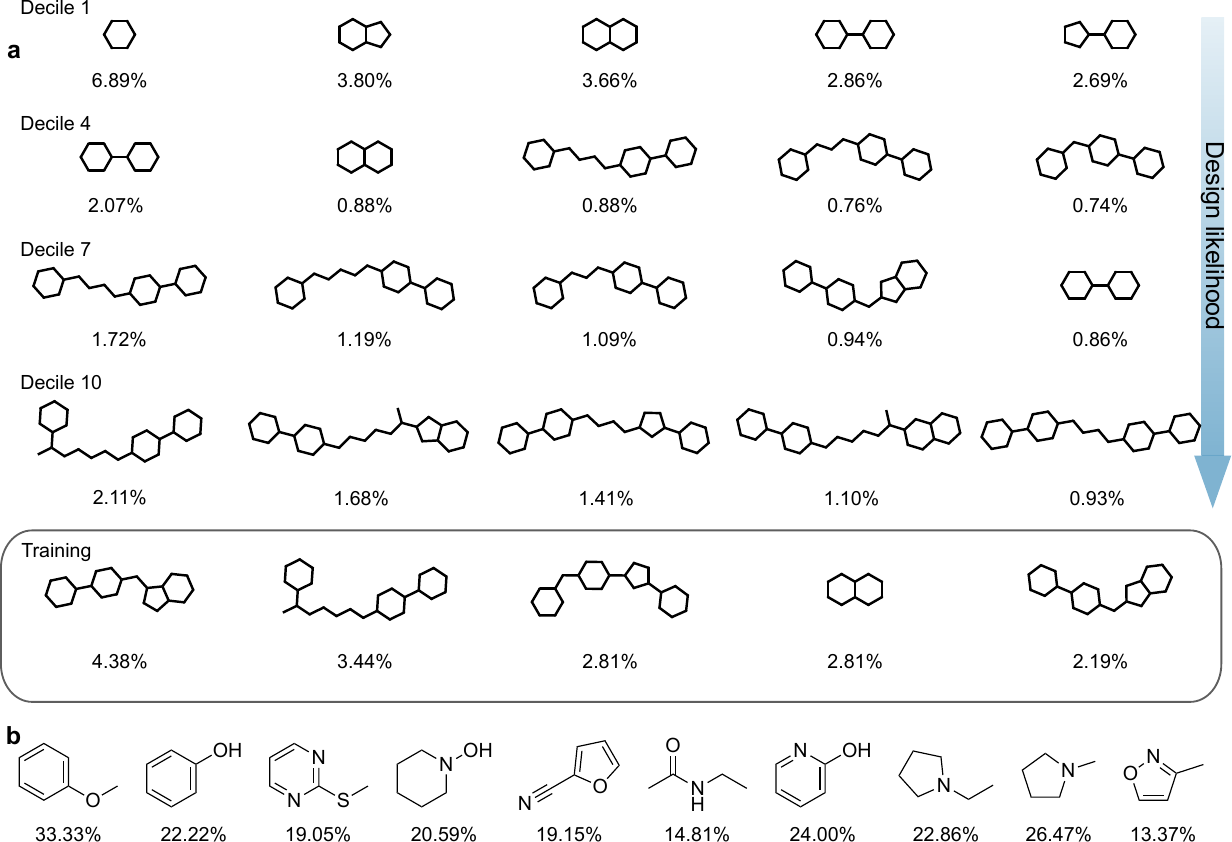}
    \caption{\textit{Likelihoods and model hallucinations.}
     Designs of LSTM trained on a DRD3 dataset are binned into increasing likelihood deciles. \textbf{(a)} The most repeating generic Bemis-Murcko scaffold is visualized for deciles 1, 4, 7, and 10, as well as the training set. The number below each scaffold denotes its frequency in the library. \textbf{(b)} Highly frequent (sampled more than ten times) and least likely designs. Maximum structural similarity to the fine-tuning sets is reported (Tanimoto similarity on extended connectivity fingerprints) below.
     }
    \label{fig:scaffold_analysis}
\end{figure*}

\subsection*{Selecting the most likely and frequent generations might hinder prospective studies}
Large molecule libraries help overcome the observed `size trap' and increase the chances of hit-finding\mbox{\cite{liu2025impact,segler2018generating}}.
However, this comes at increased computational costs when ranking and selecting molecules from large libraries. Often, criteria based on subjective judgment or expertise are used, making the analysis prone to bias and limiting its broader applicability. Recently, model likelihoods have been suggested as a model-dependent strategy (Equation (\ref{eq:likelihood})) to capture how well a SMILES sequence aligns with the information learned by the model during training \cite{moret2022perplexity, ozccelik2024chemical}. Likelihoods can be computed for any design for models trained with maximum likelihood estimation, \eg autoregressive models, variational auto encoders, and normalizing flows\cite{kingma2022autoencodingvariationalbayes,papamakarios2021normalizing,hochreiter1997long,rezende2015variational}, and external scores  (\eg discriminator predictions in adversarial settings) might be used as a replacement otherwise\cite{goodfellow2014generative}. 
Similarly, design frequencies have been recently used for molecule prioritization in prospective studies, under the hypothesis that frequently generated molecules might indicate relevant designs\mbox{\cite{ballarotto2023lowdata,isigkeit2024automated}}
. While likelihoods and design frequencies allow bypassing the need for external ranking tools, an open question remains as to how they can be used systematically to navigate large design libraries. Here, we investigate these two metrics for library prioritization, and for the information they provide on the selected designs.

\underline{Likelihood}. 
After generating 1,000,000 designs from a fine-tuned model (LSTM in this selected example), we computed the designs' likelihoods and binned them into deciles of increasing likelihood. 
We inspected the designs of each decile for: (i) syntactic score, \ie the fraction of chemically valid SMILES strings (validity) and molecules not in the training sets (novelty); (ii) structural similarity to the fine-tuning set, computed as maximum Tanimoto similarity on extended connectivity fingerprints \cite{rogers2010extended} of novel and unique designs; and (iii) number of substructures, to capture the internal diversity of each decile.

\begin{figure*}
    \centering
    \includegraphics[width=\textwidth]{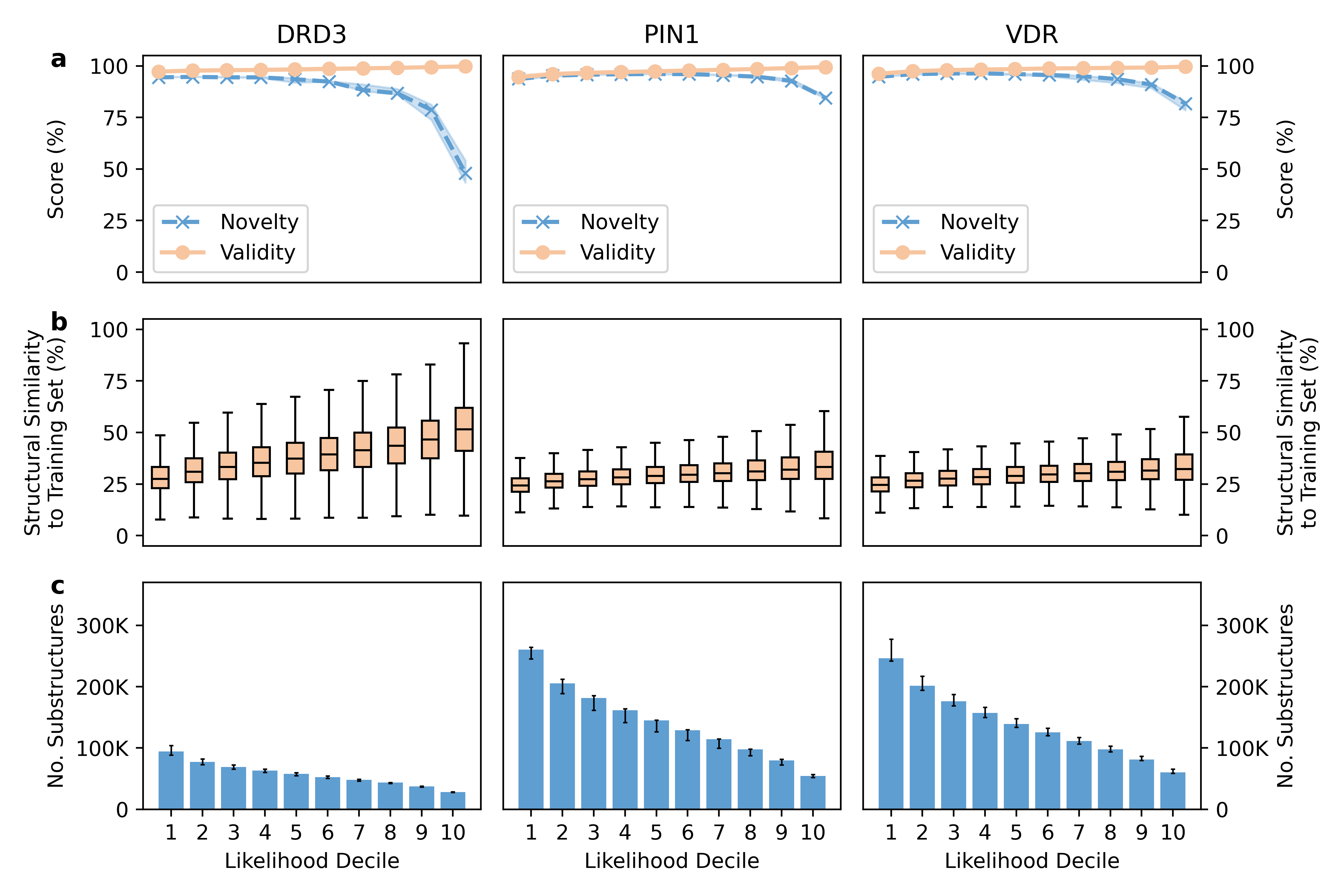}
    \caption{\textit{Navigating the design libraries of S4 architecture.} The designs of fine-tuned S4 models are divided into smaller libraries of increasing design likelihoods. Validity \textbf{(a)}, novelty \textbf{(a)}, structural similarity to the training set \textbf{(b)}, and internal diversity \textbf{(c)} are visualized as in Figure \ref{fig:explore_vs_exploit}.}
    \label{fig:explore_exploit_s4}
\end{figure*}

\begin{figure*}
    \centering
    \includegraphics[width=\textwidth]{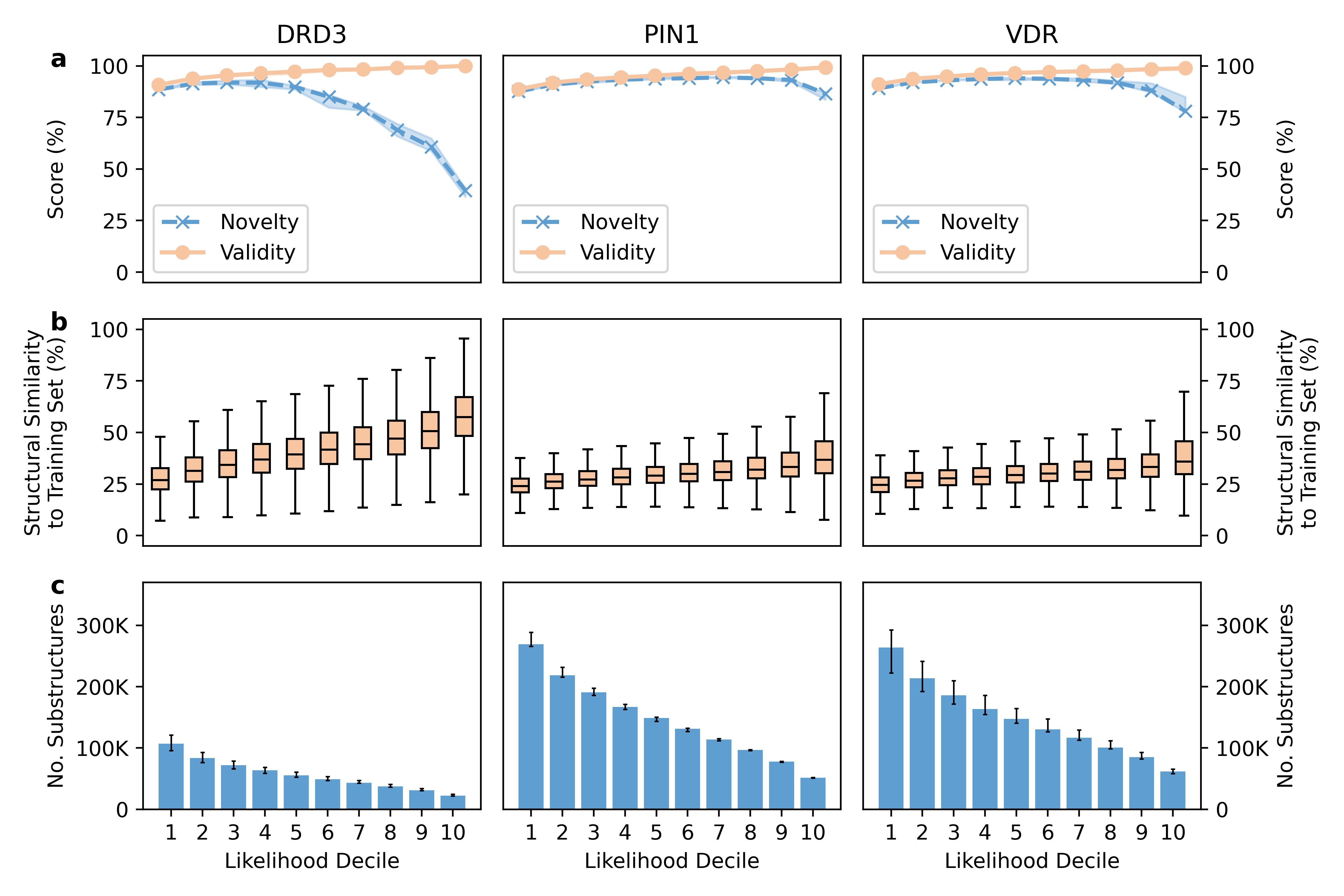}
    \caption{\textit{Navigating the design libraries of GPT architecture.} The designs of fine-tuned GPT models are divided into smaller libraries of increasing design likelihoods. Validity \textbf{(a)}, novelty \textbf{(a)}, structural similarity to the training set \textbf{(b)}, and internal diversity \textbf{(c)} are visualized as in Figure \ref{fig:explore_vs_exploit}.}
    \label{fig:explore_exploit_gpt}
\end{figure*}

The likelihood deciles revealed an exploration-exploitation trade-off. Higher likelihood bins show higher validity and structural similarity to active molecules (exploitation) \figref{fig:explore_vs_exploit}{a,b},  but contain fewer novel molecules and substructures \figref{fig:explore_vs_exploit}{a,c}.
In contrast, decreasing likelihoods favor exploration (generating novel molecules and substructures) at the cost of similarity to known bioactives and validity. Validity decreases for extremely low likelihood values -- potentially indicating problematic molecules\mbox{\cite{skinnider2024invalid}}. 
These trends are consistent across model architectures (Fig. \mbox{\ref{fig:explore_exploit_s4}}, \mbox{\ref{fig:explore_exploit_gpt}}) and targets. 

We then analyzed the most frequently occurring generic Bemis-Murcko scaffolds\cite{bemis1996properties} among the designs across likelihood deciles (1$^{st}$, 4$^{th}$, 7$^{th}$, and 10$^{th}$), in comparison with the respective fine-tuning sets (Fig. \ref{fig:scaffold_analysis}, on DRD3). Bins with higher likelihood featured frequent scaffolds identical or similar to those of active molecules, while lower bins contained simpler, repeated scaffolds, such as single or fused rings \figref{fig:scaffold_analysis}{a}. Overall, these observations show that, while likelihoods can aid the navigation of design libraries based on the envisioned application (\eg chemical space exploration vs. hit-to-lead optimization), selecting designs with extreme likelihood values has a detrimental effect.

\begin{figure*}
    \centering
    \includegraphics[width=\textwidth]{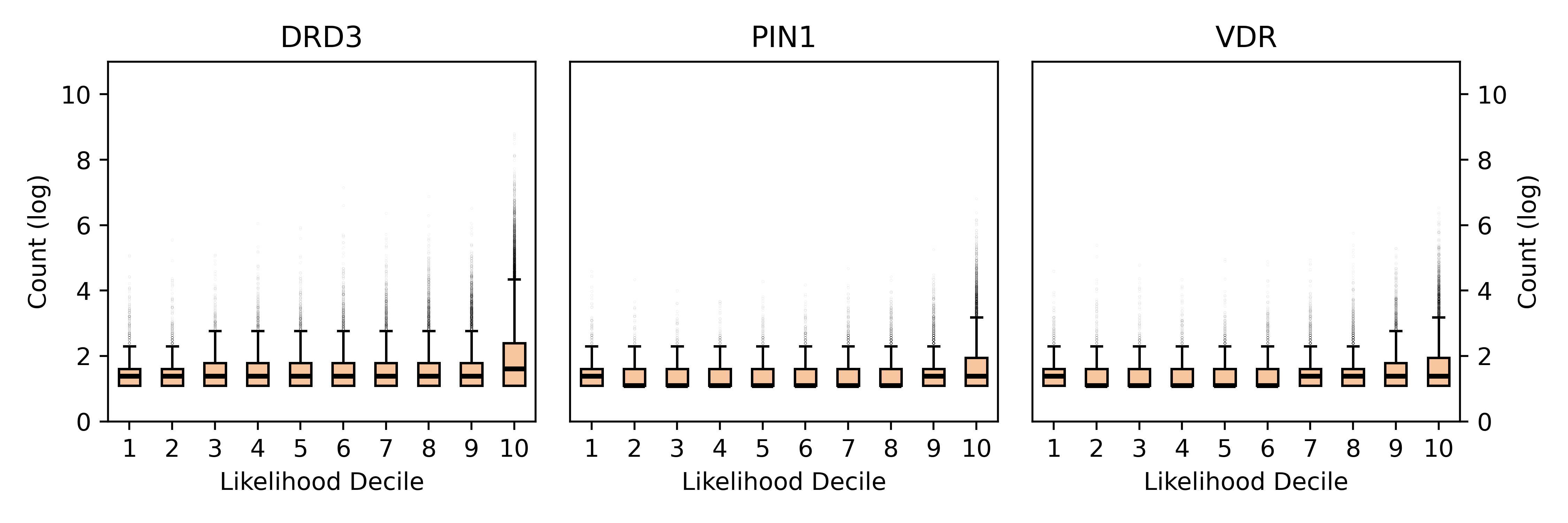}
    \caption{\textit{Design likeligood and frequency.} The repeated designs of the LSTM models per target are divided into increasing likelihood bins and their log-frequencies are visualized as a box plot.}
    \label{fig:ll_vs_count}
\end{figure*}

\underline{Design frequency}. 
When analyzing the frequently occurring molecular structures (generated more than ten times), we found that frequent designs can have low quality, consisting only of simple substructures (e.g., benzene, amine, and ether groups), making them unsuitable for prospective studies \figref{fig:scaffold_analysis}{b}. These low-quality designs, similar to `recurring hallucinations' in language models \cite{huang2023survey}, appear in the least likely decile, despite being frequently generated \figref{fig:ll_vs_count}{\text{}}. This `count trap' underscores the need to integrate likelihoods into frequency-based evaluations to avoid overemphasizing such designs.

Ultimately, our analysis reveals that relying on frequency-based ranking can lead to the selection of low-quality designs if not combined with additional evaluation strategies. Model likelihood emerged as a cost-efficient, model-intrinsic complementary metric that helps identify and filter out low-quality, repetitive generations -- akin to recurring hallucinations.

\subsection*{Chemical vocabulary size constrains structural diversity}

\begin{figure*}
    \centering
    \includegraphics[width=\textwidth]{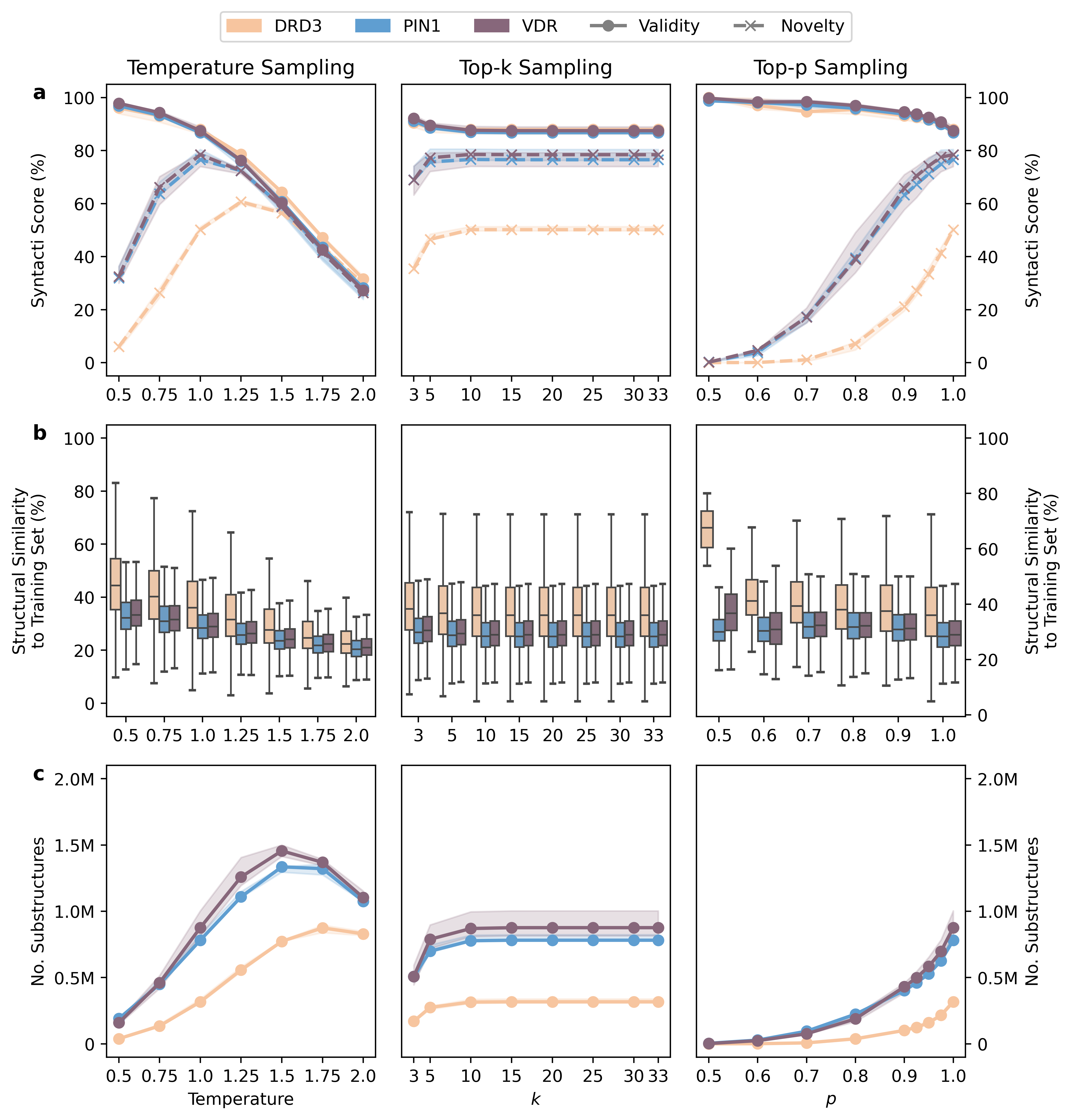}
    \caption{\textit{Benchmarking molecule sampling strategies.} The fine-tuned LSTM models across datasets are sampled using temperature, top-$k$, and top-$p$ sampling, at different temperatures, $k$, and $p$ values. 1,000,000 designs are produced per dataset split and sampling parameter combination. \textbf{(a)} Syntactic quality of the designs as measured by the fraction of valid (validity) and unique and novel compounds (novelty). The lines denote the median across five repetitions and the borders of the shaded areas display first and third quartiles. \textbf{(b)} Maximum structural similarity of each design to the respective training set is computed (as Tanimoto similarity on extended connectivity fingerprints \cite{rogers2010extended}) and the values across dataset splits are visualized as boxplots  (n$\approx$ 5,000,000). \textbf{(c)} Diversity of the designs is measured via the number of structures, \ie the number of unique Morgan keys identified \cite{rogers2010extended}. The lines denote the median of five runs and the shaded regions denote the inter-quartile ranges.}
    \label{fig:sampling_strategies}
\end{figure*}
\begin{figure*}[t]
    \centering
    \includegraphics[width=0.85\textwidth]{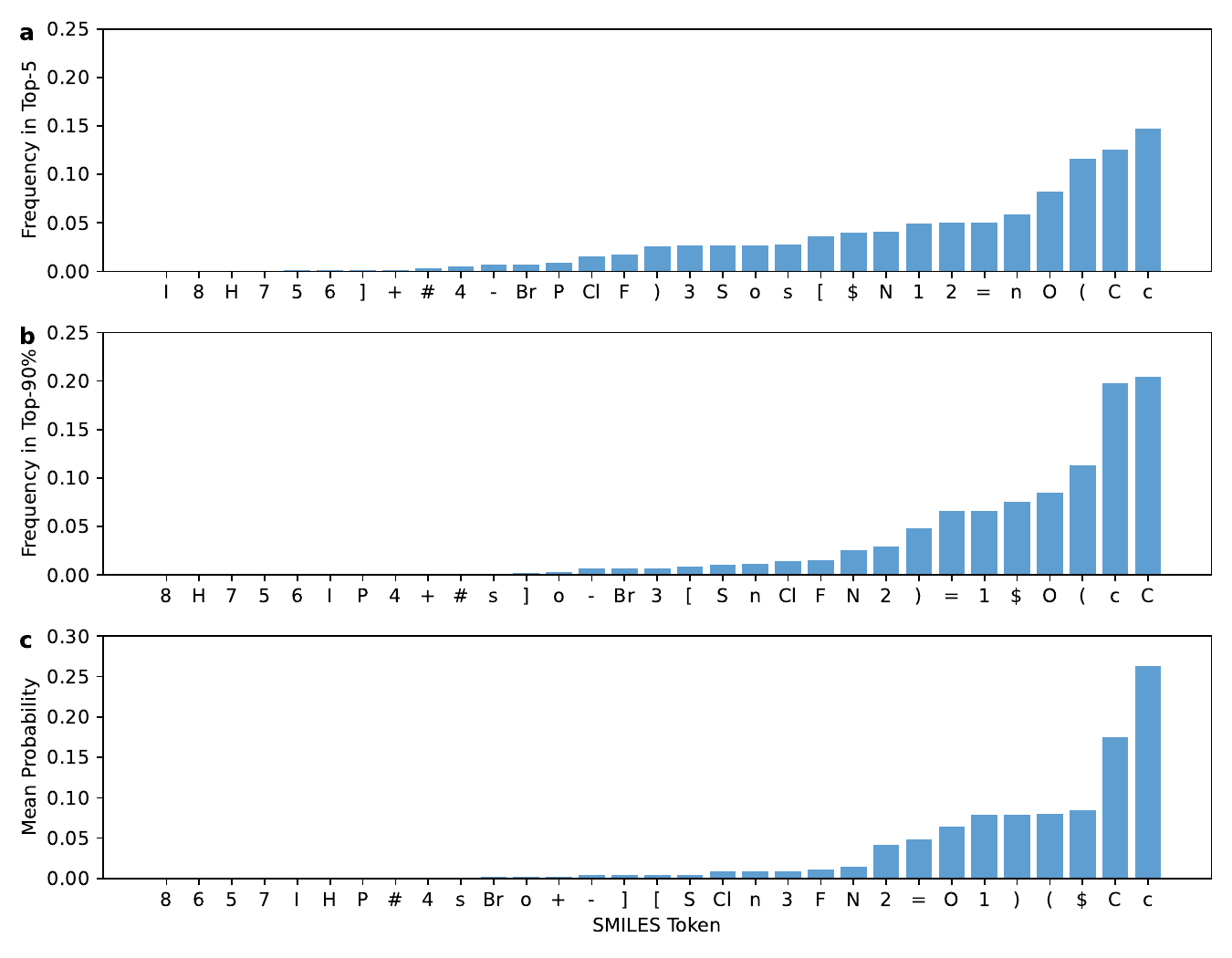}
    \caption{\textit{The curious case of molecule sampling}. 102,400 designs are generated with an LSTM model fine-tuned on the VDR dataset. The frequency of appearing in \textbf{(a)} top-5, \textbf{(b)} top-90\% of the distribution, and \textbf{(c)} mean sampling probability across generation steps is computed per SMILES token. Element symbols annotate the atom types in SMILES strings (lower-casing corresponds to aromaticity), whereas bonds are encoded with `='  (double bond) and `\#' (triple bond) tokens \cite{weininger1988smiles}. Opening and closing brackets denote branch beginning and ending, respectively, and digits define ring structures. Square brackets, `+', and `-' signs are used for explicit charge annotations. The '\$' symbol is used to denote the end of the SMILES string for this plot.}
    \label{fig:token_probs}
\end{figure*}

Another key step in the evaluation of generative drug discovery approaches is the generation of molecules themselves. This step involves sampling from the probability distributions learned by the model, as the full distribution is not directly accessible. Temperature sampling -- based on weighted random sampling of tokens (Equation (\ref{eq:t_sampling})) -- is the most common sampling approach to date for de novo design\cite{polykovskiy2020moses,brown2019guacamol}. However, when looking at the field of natural language processing, two other strategies have shown better performance\cite{holtzman2019curious,fan2018hierarchical}: (a) top-\textit{k} sampling \cite{fan2018hierarchical,zhang2020trading}, which considers only the most likely $k$ tokens at each generation step, and (b) top-\textit{p} sampling \cite{holtzman2019curious}, which uses only the most likely token subset that covers more than $p$\% of the probability distribution.  Although these strategies outperform temperature sampling in natural language processing, they have found limited application in the molecular domain \cite{moret2023leveraging}.

Using the fine-tuned LSTM models, we used each sampling strategy to generate 1,000,000 designs, using different values of temperature $T$ ($0.5 \leq T \leq 2$), $k$ ($3 \leq k \leq 33$), and $p$ ($0.5 \leq k \leq 1$). Increasing $T$, $k$, and $p$ increases the randomness of sampling (\textit{see} Methods). We measured validity, novelty, structural similarity to the fine-tuning set, and number of substructures.

\begin{figure*}
    \centering
    \includegraphics[width=\textwidth]{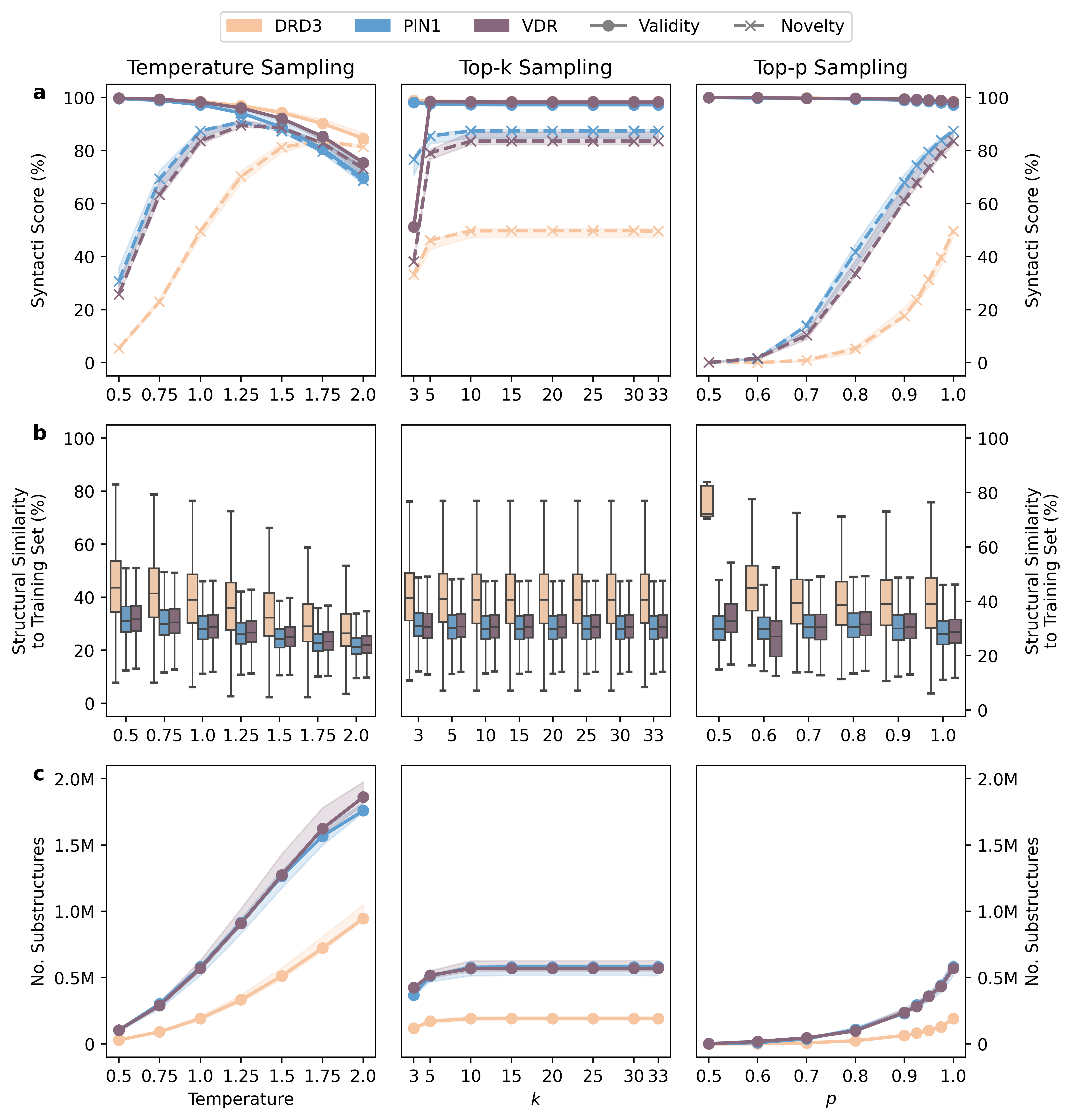}
    \caption{\textit{Benchmarking molecule sampling strategies with S4.} Temperature, top-$k$, and top-$p$ sampling are used to generate molecules with the fine-tuned S4 models, in increasing parameters. Syntactic quality \textbf{(a)}, structural similarity to the fine-tuning set \textbf{(b)}, and internal diversity \textbf{(c)} are visualized. Same sampling and plotting parameters are used as Figure \ref{fig:sampling_strategies}.}
    \label{fig:sampling_strategies_s4}
\end{figure*}

\begin{figure*}
    \centering
    \includegraphics[width=\textwidth]{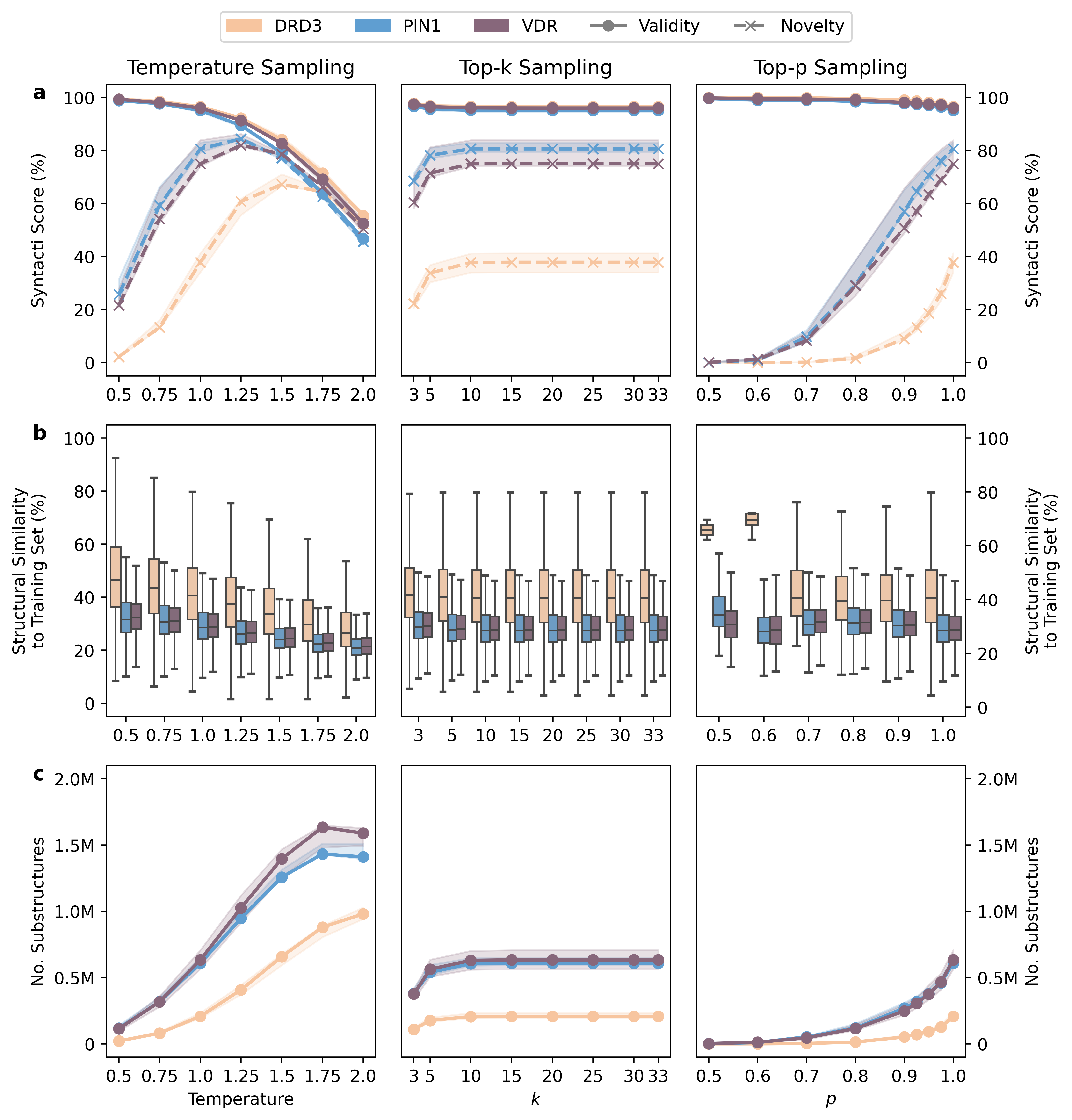}
    \caption{\textit{Benchmarking molecule sampling strategies with GPT.} Temperature, top-$k$, and top-$p$ sampling are used to generate molecules with the fine-tuned GPT models, in increasing parameters. Syntactic quality \textbf{(a)}, structural similarity to the fine-tuning set \textbf{(b)}, and internal diversity \textbf{(c)} are visualized. Same sampling and plotting parameters are used as Figure \ref{fig:sampling_strategies}.}
    \label{fig:sampling_strategies_gpt}
\end{figure*}

\begin{figure*}
    \centering
    \includegraphics[width=\textwidth]{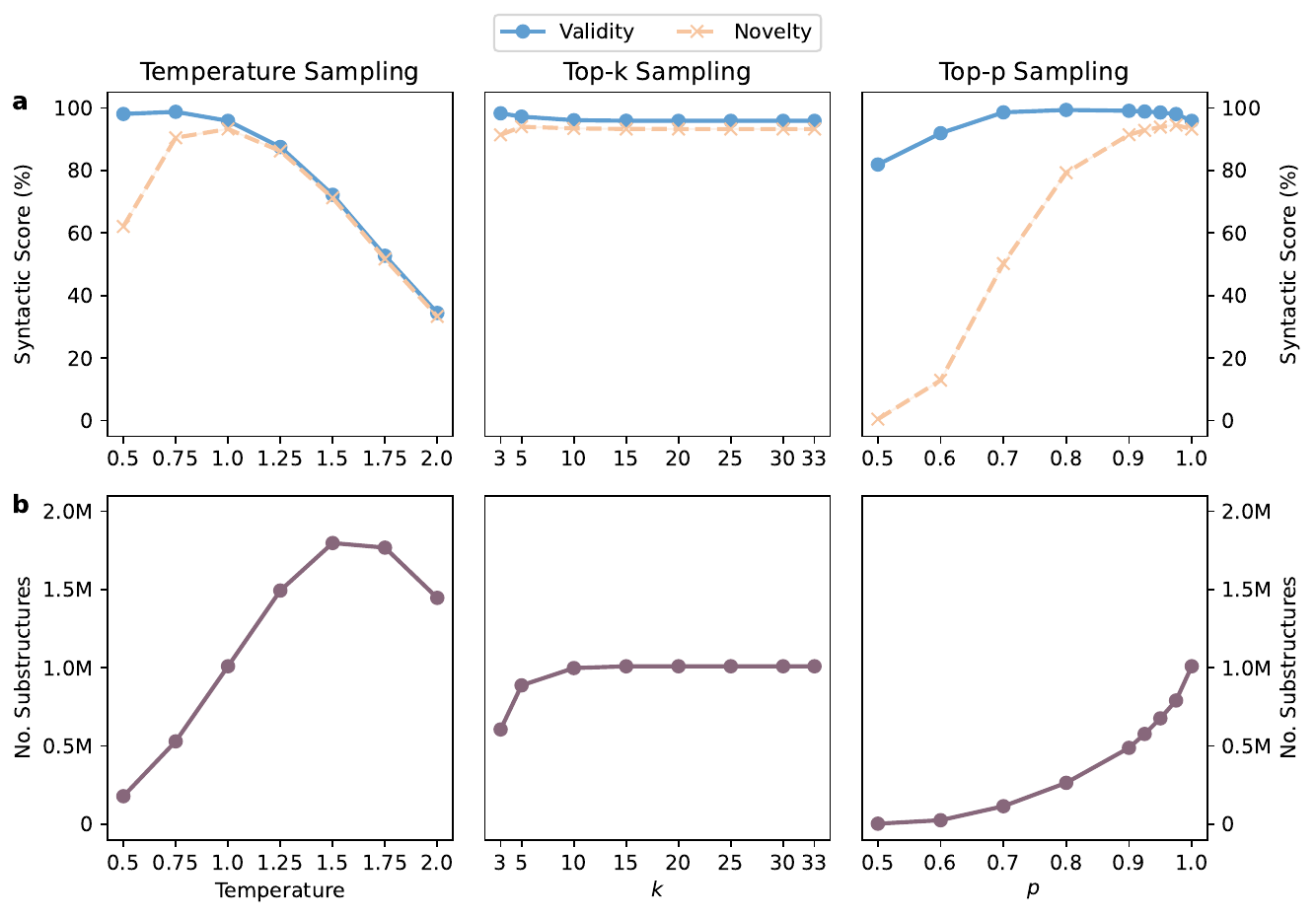}
    \caption{\textit{Benchmarking molecule sampling strategies with pretrained LSTM.} Temperature, top-$k$, and top-$p$ sampling are used to generate molecules with the pretrained models, in increasing parameters. Syntactic quality \textbf{(a)} and internal diversity \textbf{(b)} are visualized. Same sampling and plotting parameters are used as Figure \ref{fig:sampling_strategies}.}
    \label{fig:sampling_strategies_pt_lstm}
\end{figure*}

Temperature has the most substantial impact on the characteristics of de novo designs \figref{fig:sampling_strategies}{\text{}}. Higher temperature values increase the design diversity across datasets \figref{fig:sampling_strategies}{a,c}, but reduce validity and similarity to the fine-tuning set \figref{fig:sampling_strategies}{a,b}, in agreement with previous works\cite{moret2023leveraging,ozccelik2024chemical}.
For top-$k$ sampling, the smallest value of \textit{k} ($k=3$) causes mode collapse, \ie designs are valid but repetitive \figref{fig:sampling_strategies}{a}, suggesting that considering three candidate tokens might be insufficient to design diverse molecules. For $k\geq5$, the sampling behavior approximates temperature sampling with $T=1$, indicating that the top-5 tokens already cover the most likely tokens. Top-$p$ sampling behaves similarly to top-k sampling. For $p\leq0.9$, designs are valid but repetitive, while larger values of $p$ resemble temperature sampling ($T=1$). In the former case, few tokens cover the $p$ threshold, causing mode collapse; in the latter case, adding more tokens has little effect due to their low probability. We call this behavior, which is consistent across model architectures and training regimes (Fig. \ref{fig:sampling_strategies_s4}, \ref{fig:sampling_strategies_gpt}, \ref{fig:sampling_strategies_pt_lstm}),  `filtering trap'.

These findings contrast with natural language generation: while top-$k$ and top-$p$ outperform temperature sampling to generate natural language, temperature sampling is the primary sampling strategy to control molecule diversity. To gain further insights, we generated 102,400 designs (using the fine-tuned LSTM for VDR). We analyzed the frequency of appearance of each token in the top-5 and top-90\% of the distribution \figref{fig:token_probs}{a,b}, along with their mean sampling probability \figref{fig:token_probs}{c}. While probable tokens, such as carbon, are consistently likely across generation steps, most tokens rarely appear among the top-5 and top-90\%. This skewed distribution differentiates molecule design from natural language generation, where tokens can be sampled among hundreds of thousands of candidates. In the molecular field, the number of tokens is inherently constrained due to the number of elements that can be used in drug-like molecules, leading to a higher concentration of likely candidates and a less diverse set of options at each generation step. Furthermore, training chemical language models enforces the validity of generations, \eg every branch and ring opening has to be closed, further limiting the options at parts of the generation step. Overall, our analysis reveals a unique behavior of molecule generation, and exemplifies the gaps between natural language and drug discovery applications.

\begin{table*}
    \centering
    \caption{\textit{Summary of identified pitfalls, solutions, and guidelines.} These considerations were divided based on the evaluation stage they pertain to.}
    \begin{tabular}
    {p{.095\linewidth}p{.25\linewidth}p{.25\linewidth}p{.3\linewidth}}
    \toprule
    \textbf{Stage} & \textbf{Pitfalls} & \textbf{Solutions} & \textbf{Recommendations} \\
    
    \midrule
    \textbf{Library similarity} & \textbf{Library size}. 
    
    Metrics like FCD and FDD are dependent on library size, and decrease with increasing number of designs. & \textbf{Scaling up}. 
    
    Similarity metrics plateau for large libraries, making them suitable even with few reference bioactive molecules. & \textbf{Evaluating large libraries.}
    
    Report FCD and FDD values for more than 100,000 designs. \\  \midrule
    
    \textbf{Internal diversity} &\textbf{Uniqueness artifacts.} 
    
    Uniqueness decreases as library size increases and can rank models differently at different scales. & \textbf{Number of substructures.}
    
    The number of substructures is a compute-efficient and size-invariant measure of internal diversity. & \textbf{Size-invariant metrics.}
    
    The number of clusters and the number of substructures provide consistent rankings and highlight model differences when large library sizes ($\ge 10^5$) are considered. \\ \midrule
    
    \textbf{Molecule selection} &  \textbf{Excessive likelihood.} 
    
    Highly-likely designs favor exploitation (similarity to known actives) but sacrifice novelty and diversity, limiting chemical space exploration.
    
    & \textbf{Likelihood binning.}
    
    Likelihood deciles enable systematic library analysis and trade-off tuning for exploration vs. exploitation, tailored to study goals. & \textbf{Likelihood tuning.}
    
    Select likelihood deciles for specific objectives: favor exploration for hit identification or exploitation for lead optimization. \\
    
   & \textbf{Count trap.} 
   
   Models over-generate simple and repetitive substructures, resulting in low-quality designs unsuitable for follow-up studies.
    & 
    \textbf{Likelihood binning.
    }
    
    Likelihood-based scoring uncovers repetitive, low-value generations and connects generative drug discovery to NLP phenomena like ``recurring hallucinations". & 
    
    \textbf{Likelihood-guided filtering.}
    
    Use likelihood and structural evaluations jointly to identify and de-prioritize frequent but poor-quality designs in library analysis.\\

     \midrule
    
    \textbf{Molecule generation} & \textbf{Token filtering.} 
    
    Considering a small token subset during molecule sampling (via top-k or top-p sampling) cause mode collapse (repetitive, low-diversity designs). & \textbf{Temperature sampling}.
    
    Temperature sampling is the most effective strategy to control diversity and balance novelty vs. validity. & \textbf{Varying \textit{T} values. 
    }
    
    Controlling diversity through token subsets is ineffective due to the unique behavior of molecule sampling. Temperature sampling should be used, with varying $T$ to tune lead optimization and diversity-focused exploration.\\
    \bottomrule
\end{tabular}
\label{tab:traps_and_treasures}
\end{table*}

\section*{Conclusions}

Robust evaluation pipelines are essential to identify and expand the boundaries of generative deep learning in drug discovery. Despite the topic of model benchmarking having garnered remarkable attention \cite{brown2019guacamol,polykovskiy2020moses}, to date, no standardized guidelines exist on what choices to make when evaluating generative models and their designs. Our large-scale analysis across targets, metrics, and generative deep learning approaches uncovered previously overlooked factors of the evaluation pipeline that can distort the outcome of generative deep learning projects \mbox{\tabref{tab:traps_and_treasures}}.

One key finding of this study is the confounding effect of the number of generated designs on model quality evaluation. This issue has significant implications, as it can lead to an over- or underestimation of relative model performance (and design quality). To mitigate this, metrics of similarity, internal diversity, and distribution-learning capabilities should always be compared across libraries of the same size, regardless of model setup or architecture. To ensure a robust assessment of generative models, we recommend reporting these metrics for libraries containing at least 10$^5$ designs. Additionally, whenever possible, analyzing how chosen metrics vary with library size could further highlight potential pitfalls in current evaluation practices.

This required library size is significantly larger than those used in most existing benchmarks and comparative studies, highlighting the need for a re-evaluation of model performance in light of these findings -- particularly when comparing de novo designs from different studies. Moreover, generating and evaluating such large-scale de novo designs calls for more cost-efficient assessment strategies. In this work, we have identified computationally efficient measures of distributional similarity (FDD) and internal diversity (number of substructures) as one of those examples. 

When exploring model-centric approaches for ranking de novo designs, model likelihood emerged as a strategy to balance exploration and exploitation. However, extreme likelihood values -- either too high or too low -- often correspond to redundant or low-quality designs. Our analyses further revealed that specific low-quality molecules tend to appear frequently, underscoring the need for effective filtering strategies. In this regard, combining model likelihood with design frequency provides a promising, model-informed ranking approach. Nevertheless, it remains unclear how model likelihood correlates with more complex molecular properties -- such as bioactivity and toxicity -- beyond simple measures of molecular similarity. To address this, we encourage the community to systematically report model likelihoods or their analogs for each selected design, helping to `illuminate the opaque box' by clarifying what these likelihoods capture and how they can be more effectively leveraged.

This study also draws distinctions between generating `language of chemistry' and natural language. Unlike natural language, where tokens can be chosen from a vast vocabulary, molecular generation is inherently constrained by the limited number of chemical elements and feasible substructures. As a result, model predictions tend to be more concentrated around a narrower set of high-likelihood candidates, leading to different challenges in ensuring diversity and exploration. In this context, an interesting direction is fragment-based molecular representations \cite{noutahi2024safe,mastrolorito2025fragsmiles,cheng2023groupselfies}, which increases the number of available tokens. This 'chemical word' level representation (different from the atom-level representation that is routinely used for de novo design) might help strengthen the bridges between natural and chemical language processing. It remains to be determined whether this change would reflect in an increased benefit of different sampling strategies, an area that warrants further investigation.

Overall, we discovered ‘traps’ and solutions to evaluate generative drug discovery approaches. While we focused on fine-tuned chemical language models, our results are expected to be applicable to evaluate a variety of generative deep learning approaches, \eg graph-based approaches or goal-directed design \cite{popova2018deep,abate2023graph,brown2019guacamol}, and stimulate further research on potential caveats on the evaluation of generative drug discovery approaches. Meanwhile, we expect this work to set new standards to evaluate and compare different generative approaches for drug discovery, as well as novel tools for practitioners to generate promising molecular libraries and effectively navigate them in search of novel bioactive matter.

\section*{Methods}

\subsection*{Datasets}
\underline{Pre-training set}. 2,372,675 SMILES strings were obtained from ChEMBL v33 \cite{gaulton2017chembl}. Salts were removed, and molecules composed only of C, H, O, N, S, P, F, Cl, Br, and I atoms were retained. The SMILES strings of the remaining compounds were sanitized, canonicalized, and the charge and stereochemistry annotations were removed. SMILES strings longer than 80 tokens were dropped. The final set consisting of 1,584,858 molecules was randomly divided into training ($n$ = 1,500,000), validation ($n $= 40,000), and test splits ($n$ = 44,858).

\underline{Fine-tuning sets}. The fine-tuning datasets were curated from ExCAPE-DB \cite{sun2017excape}. The targets Dopamine Receptor D3 (DRD3),  peptidyl-prolyl cis/trans isomerase (PIN1), and Vitamin D Receptor (VDR) were selected. The bioactive molecules available in the pre-training set were excluded, and the remaining molecules underwent the same pre-processing steps as described above. 320 training, 128 validation, and 128 test molecules were randomly sampled among the pre-processed strings. Random sampling was repeated five times with different random seeds, obtaining five fine-tuning sets per protein target.

\subsection*{Model training}
A hyperparameter search was performed for model pre-training. 100 random hyperparameter combinations were sampled for each model (from a grid of 500, 405, 960 possible combinations for LSTM, S4, and GPT, respectively, Table \ref{tab:hp_spaces}). With all trained models, 8192 designs were generated and the hyperparameter combination whose designs yielded the highest novelty was chosen for follow-up fine-tuning. Early stopping on validation loss (cross-entropy) was used with a patience of five epochs for pre-training and three for fine-tuning (with tolerance of $10^{-5}$).

\subsection*{Molecule sampling}
\underline{Temperature sampling.} Temperature sampling applies a smoothing parameter, temperature ($T$), to the next token logits predicted by a model. The sampling probability of each token $t$ ($p_t$) is computed as:

\begin{equation}
    p_t = \frac{\exp(y_t / T)}{\sum_t \exp(y_t / T)},
    \label{eq:t_sampling}
\end{equation}

\noindent where $T$ is the temperature parameter and $y_t$ is the logit output by the model for the token $t$. Increasing the temperature value increases the uniformity of the distribution (uniform distribution for $T \rightarrow \infty$), while decreasing the temperature value decreases the randomness (Dirac distribution for $T \rightarrow 0$). When $T$ = 1, the so-called multinomial sampling (where the model output determines the probability of generating each token) is performed. We experimented with 0.5, 0.75, 1.0, 1.25, 1.5, 1.75, and 2.0 as $T$ values in this study.

\underline{Top-$k$ sampling.} Top-$k$ sampling \cite{holtzman2019curious} samples the next token from the most likely $k$ tokens. Increasing the $k$ values to the number of tokens in the vocabulary makes it equivalent to temperature sampling. Using $k = 1$ is equivalent to greedy sampling, \ie sampling the most likely token at each step. We experimented with values of $k$ equal to 3, 5, 10, 15, 20, 25, and 30 (with a vocabulary size equal to 33).

\underline{Top-$p$ sampling.} Top-$p$ sampling \cite{holtzman2019curious} (also known as nucleus sampling) samples the next token from the minimum cardinality set whose summed probabilities exceed the threshold $p$ ($0 \leq p \leq 1$). Decreasing the value of $p$ includes fewer tokens in the selection and helps to avoid degenerate outputs, while trading diversity off \cite{holtzman2019curious}. 
Top-$p$ sampling approximates greedy sampling as $p \rightarrow 0$ and is equivalent to temperature sampling when $p = 1.0$. In this study, we experimented with values of $p$ equal to 0.5, 0.6, 0.7, 0.8, 0.9, 0.925, 0.950, and 0.975.

\underline{Sampling strategy}. From each target (3x), each data split (5x), and each model architecture (3x), we generated 1,000,000 SMILES strings per sampling strategy and sampling parameter ($T$, $k$, and $p$, 22 values tested in total). This resulted in a total of $3\times 5 \times 3 \times 22 \times 10^6 = 990,000,000 = 9.9\times 10^8 \approx 10^9$  molecules that were designed and evaluated in this study. 

\subsection*{Evaluation metrics}
\underline{Syntactic Score}. Validity was computed as the fraction of the designed SMILES strings corresponding to `chemically valid' molecules. Uniqueness was computed as the percentage of distinct canonical SMILES strings among the valid ones. Novelty was computed as the fraction of valid and unique designs that were not present in either the pre-training and fine-tuning sets.

\underline{Similarity}. 
\begin{itemize}
    \item The Fr\'echet ChemNet Distance (FCD) \cite{preuer2018frechet} was computed using the \texttt{fcd} library released by the authors. \item The Fr\'echet distance \cite{frechet1957distance} between molecular descriptor distributions (FDD) was computed using the following descriptors: octanol-wated partitioning coefficient  \cite{wildman1999prediction}, molecular weight, number of hydrogen bond donors, number of rings, and topological surface area. Molecular descriptors were computed via \texttt{rdkit}, and min-max normalized to global maximum and minimum values before distance calculation. 

\item The substructure similarity was computed via Tanimoto similarity on extended connectivity fingerprints (ECFPs) \cite{rogers2010extended}. ECFPs were computed with \texttt{rdkit} (\texttt{fpSize=2048}, and \texttt{radius=2}).

\end{itemize}

\underline{Internal diversity} The number of clusters was computed by using the sphere exclusion algorithm implemented in the \texttt{LeaderPicker} module of \texttt{rdkit}, which is equivalent to computing \#Circles metric \cite{xiemuch}. We used a distance threshold of 0.6 on the Tanimoto similarity on ECFPs. Number of substructures was calculated by counting the number of unique fingerprint keys identified by the Morgan algorithm (\texttt{rdkit}, \texttt{radius=2}).  

\subsection*{Design likelihood}
Likelihood of a design $d$ ($\mathcal{L}_d$) was computed by multiplying the sampling probability $p_t$ of each SMILES token $t$:
\begin{equation}
    \mathcal{L}_d = \prod_t p_t
    \label{eq:likelihood}
\end{equation}
\noindent where $t$ runs over the tokens in the designed sequence. The log-sum-exp trick was used to mitigate numerical instabilities and log-likelihoods for each string were divided by the number of tokens.

\section*{Availability of data and materials}

The Python code to replicate our study is on GitHub at the following URL: \hyperlink{https://github.com/molML/jungle-of-generative-drug-discovery}{https://github.com/molML/jungle-of-generative-drug-discovery}.

\section*{Competing interests}
The authors declare no competing interests.

\section*{Funding}
This research was co-funded by the European Union (ERC, ReMINDER,
101077879). Views and opinions expressed are however those of the
author(s) only and do not necessarily reflect those of the European Union
or the European Research Council. Neither the European Union nor the
granting authority can be held responsible for them. The authors also
acknowledge support from the Irene Curie Fellowship, the Centre for
Living Technologies. 

\section*{Author Contributions}
Conceptualization: both authors. Data curation: R.Ö. Formal analysis: both
authors. Investigation: both authors. Methodology: both authors. Software:
R.Ö. Visualization: R.Ö. Writing – original draft:
R.Ö. Writing – review and editing: both authors.

\section*{Acknowledgements}
The authors thank Selen Parlar and Helena Brinkmann for their feedback on the manuscript.

\bibliography{references}
\bibliographystyle{ieeetr}

\newpage
\onecolumn
\beginsupplement
\newpage
\begin{center}
    
    \Large
    Supporting Information\\
    
    \Huge
    \noindent How Evaluation Choices Distort the Outcome of Generative Drug Discovery\par
    \vspace{.5em}
    \large\noindent R{\i}za \"{O}z\c{c}elik$^{1,2}$, Francesca Grisoni$^{1,2,*}$\par

\end{center}
\noindent $^1$Institute for Complex Molecular Systems and Dept. Biomedical Engineering, Eindhoven University of Technology, Eindhoven, Netherlands.\\
\noindent $^2$Centre for Living Technologies, Alliance TU/e, WUR, UU, UMC Utrecht, Netherlands.

\newpage
\begin{table*}
\begin{center}
\caption{\textit{Values used for hyperparameter optimization.}}
\begin{small}
\begin{tabular}{p{.1\linewidth}p{.3\linewidth}p{.3\linewidth}}
\toprule
\textbf{Architecture} & \textbf{Hyperparameter} & \textbf{Values} \\
\midrule
LSTM & Number of Layers  & 1, 2, 4, 6, 8 \\
     & Model dimension & 256, 512, 1024, 2048 \\
     & Dropout & 0.0, 0.1, 0.15, 0.2, 0.25 \\ \midrule

S4 & Number of Layers  & 4, 6, 8  \\
    & Model dimension & 256, 512, 1024 \\
    & Hidden stat dim. & 128, 256, 512 \\
    & Number of SSMs & 1 \\
    & Dropout & 0.0, 0.1, 0.2 \\ \midrule

GPT & Number of Layers  & 2, 4, 6, 8 \\ 
    & Model dimension & 128, 256, 512, 1024 \\
    & Number of Attention Heads & 2, 4, 8, 16 \\
    & Dropout &  0.0, 0.1, 0.2 \\ \midrule

All & Sequence length & 82  \\ 
    & Learning rate & 1e-4, 5e-4, 1e-3, 5e-3, 1e-2 \\
    & Number of max epochs & 1000 \\
    & Batch size & 8192 \\
\bottomrule
\label{tab:hp_spaces}
\end{tabular}
\end{small}
\end{center}
\end{table*}

\end{document}